\documentclass{article}
\usepackage{spconf,amsmath,graphicx}
\usepackage{float}
\usepackage{stfloats}
\usepackage{blindtext}
\usepackage{amsfonts}
\usepackage{makecell}

\title{Multi-Subdomain Adversarial Network for Cross-Subject EEG-based Emotion Recognition}
\name{Guang Lin and Jianhai Zhang*}
\address{Hangzhou Dianzi University, Hangzhou, China}
\begin{document}
%
\maketitle
\begin{abstract}
The individual difference between subjects is significant in EEG-based emotion recognition, resulting in the difficulty of sharing the model across subjects. Previous studies use domain adaptation algorithms to minimize the global domain discrepancy while ignoring the class information, which may cause misalignment of subdomains and reduce model performance. This paper proposes a multi-subdomain adversarial network (MSAN) for cross-subject EEG-based emotion recognition. MSAN uses adversarial training to model the discrepancy in the global domain and subdomain to reduce the intra-class distance and enlarge the inter-class distance. In addition, MSAN initializes parameters through a pre-trained autoencoder to ensure the stability and convertibility of the model. The experimental results show that the accuracy of MSAN is improved by 30.02\% on the SEED dataset comparing with the nontransfer method.
\end{abstract}
\section{Introduction}
\label{Introduction}
\begin{figure}[ht]
\includegraphics[width=\linewidth]{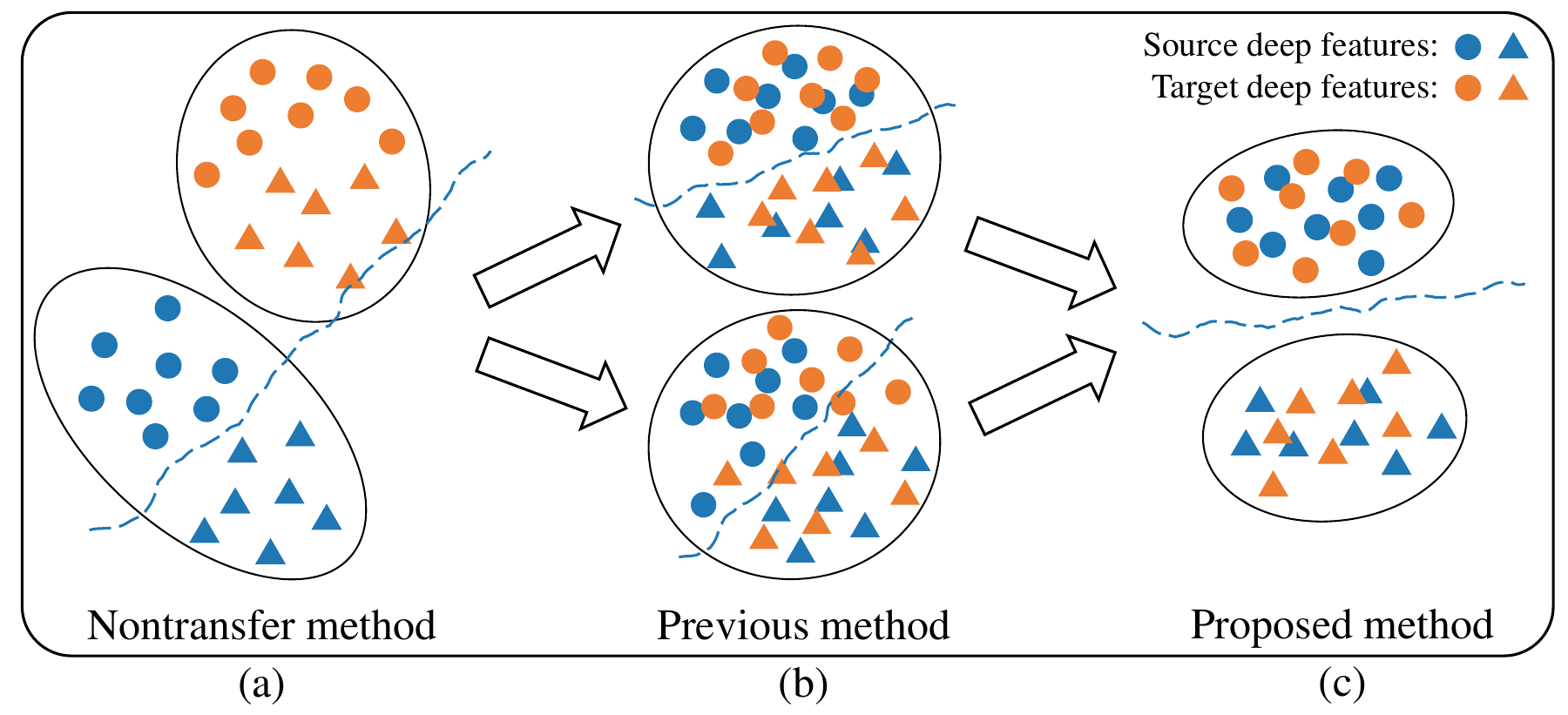}
\caption{The motivation of the proposed method. (a) The distribution of source domain data and target domain data before transferring. (b) Two kinds of distributions after global domain adaptation. (c) The distribution of the proposed method.}\label{fig:1}
\end{figure}

Emotion plays an essential role in humans' daily routines. With the development of human-computer interaction \cite{fiorini2020unsupervised} and machine learning, machines are expected to communicate with us according to human emotions, where emotion recognition is crucial \cite{cowie2001emotion,calvo2010affect}. EEG-based emotion recognition has been widely researched due to its high accuracy, low cost, and easy operation \cite{vespa1999continuous}. However, different subjects perceive emotions differently \cite{anagnostopoulos2015features}, models trained for specific subjects have poor generalization when applied to new-subject, as shown in Figure~\ref{fig:1} (a). In addition, training a specific model for each new-subject is inefficient, which requires the collection of labeled data and retrains the model \cite{shen2019challenge}. Therefore, cross-subject emotion recognition based on EEG signals is still a challenge.

In the cross-subject EEG-based emotion recognition, previous studies improved the cross-domain performance of the model by aligning source domain distribution with target domain distribution. Zheng et al. used transfer component analysis (TCA) \cite{zheng2015transfer} and transductive parameter transfer (TPT) \cite{zheng2016personalizing} to map two domains with different distributions into the same feature space and select the corresponding classifier to reduce the impact of distribution discrepancy. Li et al. \cite{li2019multisource} proposed multisource transfer learning (MSTL), which aligned the data distribution of new-subject with the multiple existing-subject separately and used the joint decision of multiple SVM classifiers to output the predicted results. Luo et al. \cite{luo2018wgan} used an adversarial domain network (DAN) with gradient reversal Layer (GRL) \cite{ganin2015unsupervised} to make the distribution of the two domains more similar. Also, based on the GRL, Zhong et al. \cite{zhong2020eeg} proposed a regularized graph neural network (RGNN) which used the graph network to extract features.

The above studies focus on minimizing the global domain discrepancy ignoring the class information of the target domain data. The misalignment of the subdomain distribution will cause poor recognition performance on the model, as shown at the bottom of Figure~\ref{fig:1} (b). In this paper, a multi-subdomain adversarial network (MSAN) is proposed for cross-subject EEG-based emotion recognition. MSAN performs both global domain adaptation and subdomain adaptation through the adversarial network. The global domain and subdomain are identified through a domain classification network. When the loss function converges, the network can align the global domain distribution while reduces the intra-class distance and enlarges the inter-class distance (Figure~\ref{fig:1} (c)), thus improving the recognition performance on the new-subject. At the same time, considering the stability of model training, an autoencoder (AE) is used to perform unsupervised training on all data, and the parameters of the first three layers of AE after training are saved as the initial parameters of MSAN.

\begin{figure*}[ht]
\includegraphics[width=\textwidth]{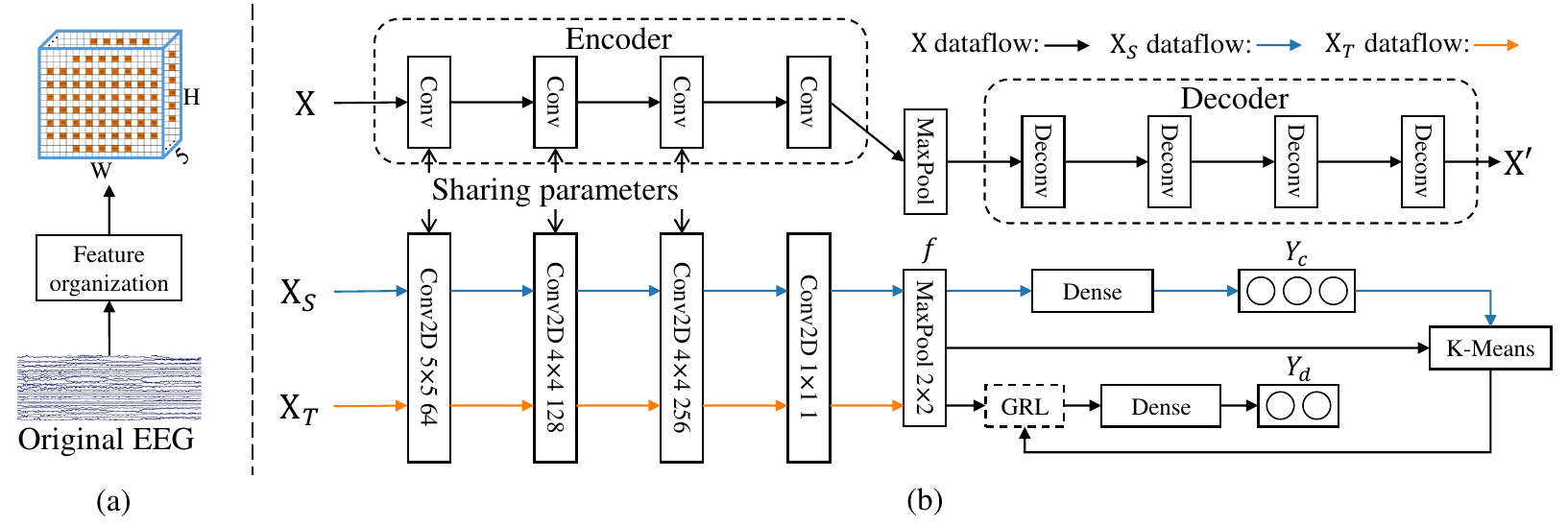}
\caption{The framework of the proposed method. (a) Feature organization. (b) MSAN training structure. All data ($X$) is input into the autoencoder for unsupervised training. Deep features ($f$) are extracted from source data ($X_S$) and target data ($X_T$) through the feature extraction network, and then are input into the two classifiers to obtain task label ($Y_c$) and domain label ($Y_d$) respectively.}
\label{fig:2}
\end{figure*}

\section{Methods}
\label{Methods}
\subsection{Feature organization}
\label{Feature organization}
Considering the factor that physiological characteristics of EEG signals vary with frequency \cite{zheng2015investigating} and encephalic regions \cite{wang2018emotionet}, this paper uses a 3D structure \cite{yang2018continuous,shen2020eeg} to organize the EEG signals, as shown in Figure~\ref{fig:2} (a). In detail, the band-pass filter is performed on the EEG signals of all channels to divide the signal into the delta, theta, alpha, beta, and gamma. Then differential entropy (DE) feature \cite{duan2013differential} in a 1-second sliding window is extracted in each frequency band. In addition, the channel dimension information is organized into a 2D map according to the electrode distribution in the data acquisition. Therefore, each segment is a 3D structure $X \in \mathbb{R}^{h \times w \times d}$, $h$ and $w$ are the width and height of the 2D map.

\subsection{Model training}
\label{Model training}
\subsubsection{Multi-subdomain adversarial network}
\label{Multi-subdomain adversarial network}
The Multi-subdomain adversarial network (MSAN) models the discrepancy in the global domain and subdomain, which reduces the intra-class distance and enlarges the inter-class distance to improve the recognition performance on the new-subject. As shown in Figure~\ref{fig:2} (b), MSAN consists of four modules: feature extraction network $G_f$, class prediction network $D_c$, domain classification network $D_d$, and K-Means clustering. The input of the network is the source domain data with task label and the target domain data without task label, as well as the domain labels. The EEG signal $X$ is first mapped into a feature through the $G_f$. The $D_c$ maps the feature of the source domain data to obtain the prediction results of task label. The $D_d$ maps the feature of all data to obtain the prediction results of domain label.

There are two constraints in the global domain adversarial: the error of the $D_c$ is minimized to achieve accurate classification of task labels of the source domain data; the error of the $D_d$ is maximized to confuse source domain data with the target domain data which minimizes the two domains discrepancy. Error maximization can be achieved through gradient reversal layer (GRL), which reverses the gradient direction during backpropagation and realizes identity during forward. The related mathematical expressions are as follows:
\begin{equation}
\mathcal{R}(x)=x, \frac{d \mathcal{R}}{d x}=-\mathrm{I}
\end{equation}
where $I$ is an identity matrix. The complete optimization objective of $G_f$, $D_c$ and $D_d$ is formulated as follows:
\begin{equation}
\begin{aligned}
\mathbb{E}\left(\theta_{f}, \theta_{c}, \theta_{d}\right) &=\mathcal{L}_{D_{c}}\left(X, \theta_{f}, \theta_{c}\right)+\mathcal{L}_{D_{d}}\left(X, \theta_{f}, \theta_{d}\right) \\
&=\sum_{n=1,d_n=0}^{N} \mathcal{L}_{c}\left(D_{c}\left(G_{f}\left(x_{n} ; \theta_{f}\right) ; \theta_{c}\right), y_{c}^{n}\right) \\
&+\sum_{n=1}^{N} \mathcal{L}_{d}\left(D_{d}\left(\mathcal{R}\left(G_{f}\left(x_{n} ; \theta_{f}\right)\right) ; \theta_{d}\right), y_{d}^{n}\right)
\end{aligned}
\end{equation}
where $X$ represents the input EEG signal, $d_n=0$ represents $x_n \in X_s$, and $\theta$ represents the corresponding network parameters. The optimal parameters are calculated by the following formula:
\begin{equation}
\left(\hat{\theta}_{f}, \hat{\theta}_{c}\right)=\arg \min _{\theta_{f}, \theta_{c}} \mathbb{E}\left(\theta_{f}, \theta_{c}, \hat{\theta}_{d}\right)
\end{equation}
\begin{equation}
\hat{\theta}_{d}=\arg \max _{\theta_{d}} \mathbb{E}\left(\hat{\theta}_{f}, \hat{\theta}_{c}, \theta_{d}\right)
\end{equation}

In the K-Means clustering \cite{yang2017towards}, we use the number of task categories in the source domain dataset as the $K$ value. Next, performing the following three steps: (1) Calculate the centroid of all source domain data points in each category and use it as the initial cluster center of the corresponding cluster. (2) Calculate the Euclidean distance between each data point in the target domain dataset and the cluster center, and assign it to the nearest cluster. (3) Recalculate the centroid of each cluster and repeat step (1) to (3) until the cluster center stabilizes. After the convergence of K-Means, the top 10\% of the target domain data points closest to the cluster center are selected and temporarily marked according to the cluster. With the class information added for subdomain adversarial, the adjusted loss function $\mathcal{L}_{D_d}$ is as follows:
\begin{equation}
\begin{aligned}
\mathcal{L}_{D_{d}}&\left(X, \theta_{f}, \theta_{d}\right) \\
&=\sum_{n=1, \ldots, N} \mathcal{L}_{d}\left(D_{d}\left(\mathcal{R}\left(G_{f}\left(x_{n} ; \theta_{f}\right)\right) ; \theta_{d}\right), y_{d}^{n}\right) \\
&+\sum_{n=1,i=j}^{0.1 * N} \mathcal{L}_{d}\left(D_{d}\left(\mathcal{R}\left(G_{f}\left(x_{n}^{\prime} ; \theta_{f}\right)\right) ; \theta_{d}\right), y_{d}^{n}\right) \\
&+\sum_{n=1,i \neq j}^{0.1 * N} \mathcal{L}_{d}\left(D_{d}\left(G_{f}\left(x_{n}^{\prime} ; \theta_{f}\right) ; \theta_{d}\right), y_{d}^{n}\right)
\end{aligned}
\end{equation}
Where $x'$ represents the data points reordered according to Euclidean distance and $i$ and $j$ represent the task labels of the samples in the two domains. When $i=j$, the domain discrepancy is reduced through GRL. When $i \neq j$, keep the gradient direction during the backpropagation to increase the domain discrepancy.

\subsubsection{Autoencoder}
\label{Autoencoder}
Since extracted feature distributions vary dramatically in the initial training stage, subdomain adaptation may cause the model to train in the wrong direction, which leads to model collapse. In order to improve the stability of MSAN, an autoencoder is applied to the pre-training \cite{erhan2009difficulty} of the model. Autoencoder \cite{wen2018deep} is a deep neural network that can learn the efficient representation of EEG signals through unsupervised learning. It consists of two main parts: encoder and decoder. The function of the encoder is to encode high-dimensional input into low-dimensional hidden variables, thereby forcing the neural network to learn the most informative features. The function of the decoder is to restore the hidden variables to the initial dimension. The autoencoder uses the data itself as supervision to guide the neural network to learn a mapping relationship. We consider the AE model defined as:
\begin{equation}
X^{\prime}=\operatorname{Decoder}(\operatorname{Encoder}(X)) \approx X
\end{equation}

\section{Dataset}
\label{Dataset}
The effectiveness of the proposed method is evaluated on the SEED dataset \cite{zheng2015investigating}. This dataset is collected by BCMI Lab. In the experiment, 15 film clips selected from the 6 Chinese films were used as the stimuli, and every 5 clips corresponded to one kind of emotion. Several types of emotions were stimulated during the experiment, including positive, negative, and neutral. Fifteen healthy subjects participated in the experiment and were asked to complete a questionnaire immediately to report the individual emotional response after watching each clip. The ESI NeuroScan System with 62-channels was used to record the EEG signals, and the sampling frequency was 1000 Hz. In order to reduce the storage space and the amount of calculation, after removing some basic noise from the data, the data was down-sampled to 200 Hz.

For SEED, the 62-channels' EEG signals is organized a 3D feature map of $H \times W \times 5$, where $H=17$, $W=19$ in the feature organization and leave-one-subject-out validation method is used to evaluate the model performance. Specifically, the data of 14 subjects are used as the training dataset, and the data of the remaining subject is used as the test dataset.

\begin{figure*}[ht]
\centering
\includegraphics[width=0.95\textwidth]{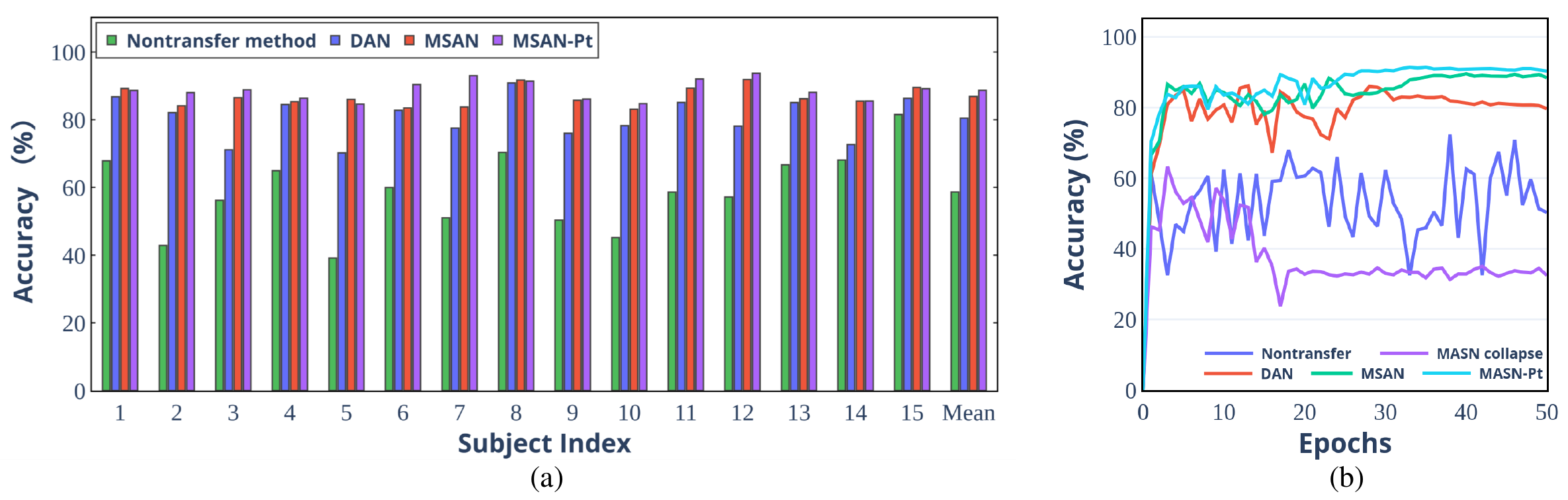}
\caption{(a) The accuracy of the four models under the cross-subject task. (b) The accuracy changes in 50 iterations.}
\label{fig:3}
\end{figure*}

\section{Results}
\label{Results}
For SEED, we compared the performance achieved by MSAN with the best results reported in the literature. The average accuracy and standard deviation (STD) represent the model performance, and all results were obtained on the data of 15 subjects. As shown in Table~\ref{table:1}, the nontransfer method was set as the baseline. MSTL took negative transfer into account and used multiple global domains to match subjects for model selection, which achieved state of the art performance. However, the STD of MSTL was large and the performance of the algorithm varies greatly on subjects. When the matching degree of the new subject and all the existing subjects was low, it was difficult to identify on the new-subject accurately. MSAN utilized multiple subdomains and reached 88.70\% accuracy rate, with minor STD and stable performance in all subjects, which proved the effectiveness of the proposed method.

\begin{table}[ht]
\centering
\caption{Means and standard deviations of multiple methods on the SEED dataset under the cross-subject task}
\label{table:1}
\begin{tabular}{l|c|c|c}
\hline \hline Model & Mean & Std. & Improve \\
\hline Nontransfer & $58.68$ & $11.21$ & - \\
\hline TCA [7] & $71.80$ & $13.99$ & $13.12$ \\
\hline TPT [8] & $76.31$ & $15.89$ & $17.63$ \\
\hline DAN [11] & $80.49$ & $6.00$ & $21.81$ \\
\hline RGNN [12] & $85.30$ & $6.72$ & $26.62$ \\
\hline WGANDA [10] & $87.07$ & $7.14$ & $28.39$ \\
\hline MSTL [9] & $88.92$ & $10.35$ & $30.24$ \\
\hline \hline Ours & $88.70$ & $2.84$ & $30.02$ \\
\hline \hline
\end{tabular}
\end{table}

In the ablation studies, we compared the nontransfer method, DAN, and MSAN to demonstrate the advantages of the proposed method. Figure~\ref{fig:3} (a) shows that the test results of the four models on the cross-subject task. In the algorithm without transfer, the average accuracy of the model was low, only 58.68\%, which was caused by individual differences. In the DAN, global domain adversarial was introduced, and its average accuracy had been significantly improved, reaching 80.49\%. Based on DAN, MSAN carried out subdomain adversarial, and compared with DAN, the average accuracy increased by 6.39\% to 86.88\%.

In order to analyze the characteristics of various models more specifically, Figure~\ref{fig:3} (b) shows the change of accuracy of subject 11 in 50 iterations. In the absence of any new-subject information, the accuracy of the nontransfer model was low and severely fluctuated. After joining the global domain adversarial in DAN, the highest accuracy was reached in the 28th iteration, but the accuracy gradually decreased in the subsequent training. At a learning rate of 0.001, the feature distribution would not change significantly after the domain discriminator network was close to convergence. However, the continuous training of the $D_c$ on the source domain data might decrease generalization ability. In MSAN, subdomain adversarial was added to align subdomains after global domain alignment to improve accuracy gradually during training. In the initial training stage, the task labels obtained were unreliable, and the misalignment would collapse the model. Therefore, we added the AE structure to pre-train MSAN (MSAN-Pt), which further improved the accuracy, but more importantly, improved the stability of MSAN.

\begin{figure}[ht]
\centering
\includegraphics[width=0.95\linewidth]{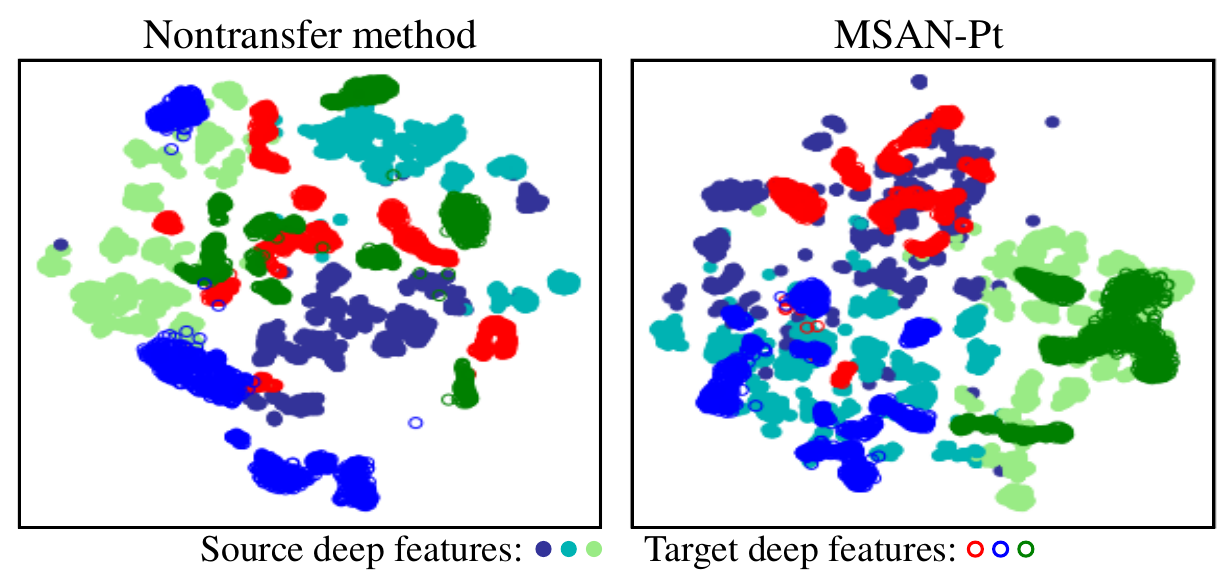}
\caption{Feature visualization.}
\label{fig:4}
\end{figure}

In order to have a better view of the effectiveness of MSAN-Pt, we used t-SNE \cite{van2008visualizing} to perform a two-dimensional visualization of the feature distributions of source domain data and target domain data, as shown in Figure~\ref{fig:4}. In the nontransfer method, the data of source subjects and target subjects had different feature distributions due to the individual differences between subjects. Therefore, when the model trained with existing-subject data was tested on the new-subject data, the classification accuracy was low. After global domain adversarial and subdomain adversarial in MSAN, the distribution of target domain data was similar to that of source domain data in the global domain and also aligned in the subdomains so that the $D_c$ correctly identified the target subjects' emotions.

\section{Conclusions}
\label{Conclusions}
In the EEG-based emotion recognition task, since the individual differences of EEG signals, the model generalization is poor applied to new-subject. This paper designs a novel framework (MSAN) based on domain adversarial network, which considers both global domain adversarial and subdomain adversarial. The average accuracy of MSAN reaches 86.88\% on the SEED dataset, which is 6.39\% higher than that of the DAN. At the same time, to improve the model stability, an autoencoder is applied to the pre-training of MSAN. The average accuracy of MSAN-Pt reaches 88.70\%, which is 30.02\% higher than that of the nontransfer method.


\bibliographystyle{IEEEbib}
\bibliography{refs}

\begin{thebibliography}{10}

\bibitem{fiorini2020unsupervised}
Laura Fiorini, Gianmaria Mancioppi, Francesco Semeraro, Hamido Fujita, and
  Filippo Cavallo,
\newblock ``Unsupervised emotional state classification through physiological
  parameters for social robotics applications,''
\newblock {\em Knowledge-Based Systems}, vol. 190, pp. 105217, 2020.

\bibitem{cowie2001emotion}
Roddy Cowie, Ellen Douglas-Cowie, Nicolas Tsapatsoulis, George Votsis, Stefanos
  Kollias, Winfried Fellenz, and John~G Taylor,
\newblock ``Emotion recognition in human-computer interaction,''
\newblock {\em IEEE Signal processing magazine}, vol. 18, no. 1, pp. 32--80,
  2001.

\bibitem{calvo2010affect}
Rafael~A Calvo and Sidney D'Mello,
\newblock ``Affect detection: An interdisciplinary review of models, methods,
  and their applications,''
\newblock {\em IEEE Transactions on affective computing}, vol. 1, no. 1, pp.
  18--37, 2010.

\bibitem{vespa1999continuous}
Paul~M Vespa, Val Nenov, and Marc~R Nuwer,
\newblock ``Continuous eeg monitoring in the intensive care unit: early
  findings and clinical efficacy,''
\newblock {\em Journal of Clinical Neurophysiology}, vol. 16, no. 1, pp. 1--13,
  1999.

\bibitem{anagnostopoulos2015features}
Christos-Nikolaos Anagnostopoulos, Theodoros Iliou, and Ioannis Giannoukos,
\newblock ``Features and classifiers for emotion recognition from speech: a
  survey from 2000 to 2011,''
\newblock {\em Artificial Intelligence Review}, vol. 43, no. 2, pp. 155--177,
  2015.

\bibitem{shen2019challenge}
Yi-Wei Shen and Yuan-Pin Lin,
\newblock ``Challenge for affective brain-computer interfaces: Non-stationary
  spatio-spectral eeg oscillations of emotional responses,''
\newblock {\em Frontiers in human neuroscience}, vol. 13, pp. 366, 2019.

\bibitem{zheng2015transfer}
Wei-Long Zheng, Yong-Qi Zhang, Jia-Yi Zhu, and Bao-Liang Lu,
\newblock ``Transfer components between subjects for eeg-based emotion
  recognition,''
\newblock in {\em 2015 international conference on affective computing and
  intelligent interaction (ACII)}. IEEE, 2015, pp. 917--922.

\bibitem{zheng2016personalizing}
Wei-Long Zheng and Bao-Liang Lu,
\newblock ``Personalizing eeg-based affective models with transfer learning,''
\newblock in {\em Proceedings of the twenty-fifth international joint
  conference on artificial intelligence}, 2016, pp. 2732--2738.

\bibitem{li2019multisource}
Jinpeng Li, Shuang Qiu, Yuan-Yuan Shen, Cheng-Lin Liu, and Huiguang He,
\newblock ``Multisource transfer learning for cross-subject eeg emotion
  recognition,''
\newblock {\em IEEE transactions on cybernetics}, vol. 50, no. 7, pp.
  3281--3293, 2019.

\bibitem{luo2018wgan}
Yun Luo, Si-Yang Zhang, Wei-Long Zheng, and Bao-Liang Lu,
\newblock ``Wgan domain adaptation for eeg-based emotion recognition,''
\newblock in {\em International Conference on Neural Information Processing}.
  Springer, 2018, pp. 275--286.

\bibitem{ganin2015unsupervised}
Yaroslav Ganin and Victor Lempitsky,
\newblock ``Unsupervised domain adaptation by backpropagation,''
\newblock in {\em International conference on machine learning}. PMLR, 2015,
  pp. 1180--1189.

\bibitem{zhong2020eeg}
Peixiang Zhong, Di~Wang, and Chunyan Miao,
\newblock ``Eeg-based emotion recognition using regularized graph neural
  networks,''
\newblock {\em IEEE Transactions on Affective Computing}, 2020.

\bibitem{zheng2015investigating}
Wei-Long Zheng and Bao-Liang Lu,
\newblock ``Investigating critical frequency bands and channels for eeg-based
  emotion recognition with deep neural networks,''
\newblock {\em IEEE Transactions on Autonomous Mental Development}, vol. 7, no.
  3, pp. 162--175, 2015.

\bibitem{wang2018emotionet}
Yi~Wang, Zhiyi Huang, Brendan McCane, and Phoebe Neo,
\newblock ``Emotionet: A 3-d convolutional neural network for eeg-based emotion
  recognition,''
\newblock in {\em 2018 International Joint Conference on Neural Networks
  (IJCNN)}. IEEE, 2018, pp. 1--7.

\bibitem{yang2018continuous}
Yilong Yang, Qingfeng Wu, Yazhen Fu, and Xiaowei Chen,
\newblock ``Continuous convolutional neural network with 3d input for eeg-based
  emotion recognition,''
\newblock in {\em International Conference on Neural Information Processing}.
  Springer, 2018, pp. 433--443.

\bibitem{shen2020eeg}
Fangyao Shen, Guojun Dai, Guang Lin, Jianhai Zhang, Wanzeng Kong, and Hong
  Zeng,
\newblock ``Eeg-based emotion recognition using 4d convolutional recurrent
  neural network,''
\newblock {\em Cognitive Neurodynamics}, vol. 14, no. 6, pp. 815--828, 2020.

\bibitem{duan2013differential}
Ruo-Nan Duan, Jia-Yi Zhu, and Bao-Liang Lu,
\newblock ``Differential entropy feature for eeg-based emotion
  classification,''
\newblock in {\em 2013 6th International IEEE/EMBS Conference on Neural
  Engineering (NER)}. IEEE, 2013, pp. 81--84.

\bibitem{yang2017towards}
Bo~Yang, Xiao Fu, Nicholas~D Sidiropoulos, and Mingyi Hong,
\newblock ``Towards k-means-friendly spaces: Simultaneous deep learning and
  clustering,''
\newblock in {\em international conference on machine learning}. PMLR, 2017,
  pp. 3861--3870.

\bibitem{erhan2009difficulty}
Dumitru Erhan, Pierre-Antoine Manzagol, Yoshua Bengio, Samy Bengio, and Pascal
  Vincent,
\newblock ``The difficulty of training deep architectures and the effect of
  unsupervised pre-training,''
\newblock in {\em Artificial Intelligence and Statistics}. PMLR, 2009, pp.
  153--160.

\bibitem{wen2018deep}
Tingxi Wen and Zhongnan Zhang,
\newblock ``Deep convolution neural network and autoencoders-based unsupervised
  feature learning of eeg signals,''
\newblock {\em IEEE Access}, vol. 6, pp. 25399--25410, 2018.

\bibitem{van2008visualizing}
Laurens Van~der Maaten and Geoffrey Hinton,
\newblock ``Visualizing data using t-sne.,''
\newblock {\em Journal of machine learning research}, vol. 9, no. 11, 2008.

\end{thebibliography}

\end{document}